\documentclass[seceq]{ptptex}

\usepackage{graphicx}
\usepackage{wrapft}

 \newcommand\beq{\begin{equation}}
 \newcommand\eeq{\end{equation}}
 \newcommand\beqn{\begin{eqnarray}}
 \newcommand\eeqn{\end{eqnarray}}

\def\fm{\,\mbox{fm}}
\def\GeV{\,\mbox{GeV}}

\def\TeV{\,\mbox{TeV}}

\def\bm{\begin{minipage}}
\def\em{\end{minipage}}
\def\bc{\begin{center}}
\def\ec{\end{center}}

\title{Boosted saturation bound in colliding nuclei}

\author{B.~Z.~Kopeliovich}

\inst{Departamento de F\'{\i}sica,
Universidad T\'ecnica Federico Santa Mar\'{\i}a;
\\
Instituto de estudios avanzados en ciencias en ingenier'a;\\
Centro Cient\'ifico-Tecnol\'ogico de Valpara\'iso;\\
Casilla 110-V, Valpara\'iso, Chile
}

\abst{Interaction with a nucleus in $pA$ collisions enhances the higher Fock components in the projectile proton.
Effectively, this corresponds to an increase of the hard scale of the process by the saturation momentum $Q^2\Rightarrow Q^2+Q_{sA}^2$, which leads to an increased gluon distribution function at small $x$ (but suppressed at $x\to1$) compared to that in $pp$ collisions.
In the case of $AA$ collisions the gluon distributions of bound nucleons in both nuclei turn out to be  enhanced, i.e. to be boosted to higher saturation scales compared to $pA$ collisions.
A set of bootstrap equations relating the saturation scales in the colliding nuclei is derived and solved~\cite{boosting}. The boosting effect has a moderate magnitude at the energies of RHIC, but becomes significant at LHC.}

\begin{document}

\maketitle

\section{Modification of the gluon PDF in pA and AA collisions}

\begin{figure}[htb]
\parbox{\halftext}{
\centerline{\includegraphics[width=4.2 cm]{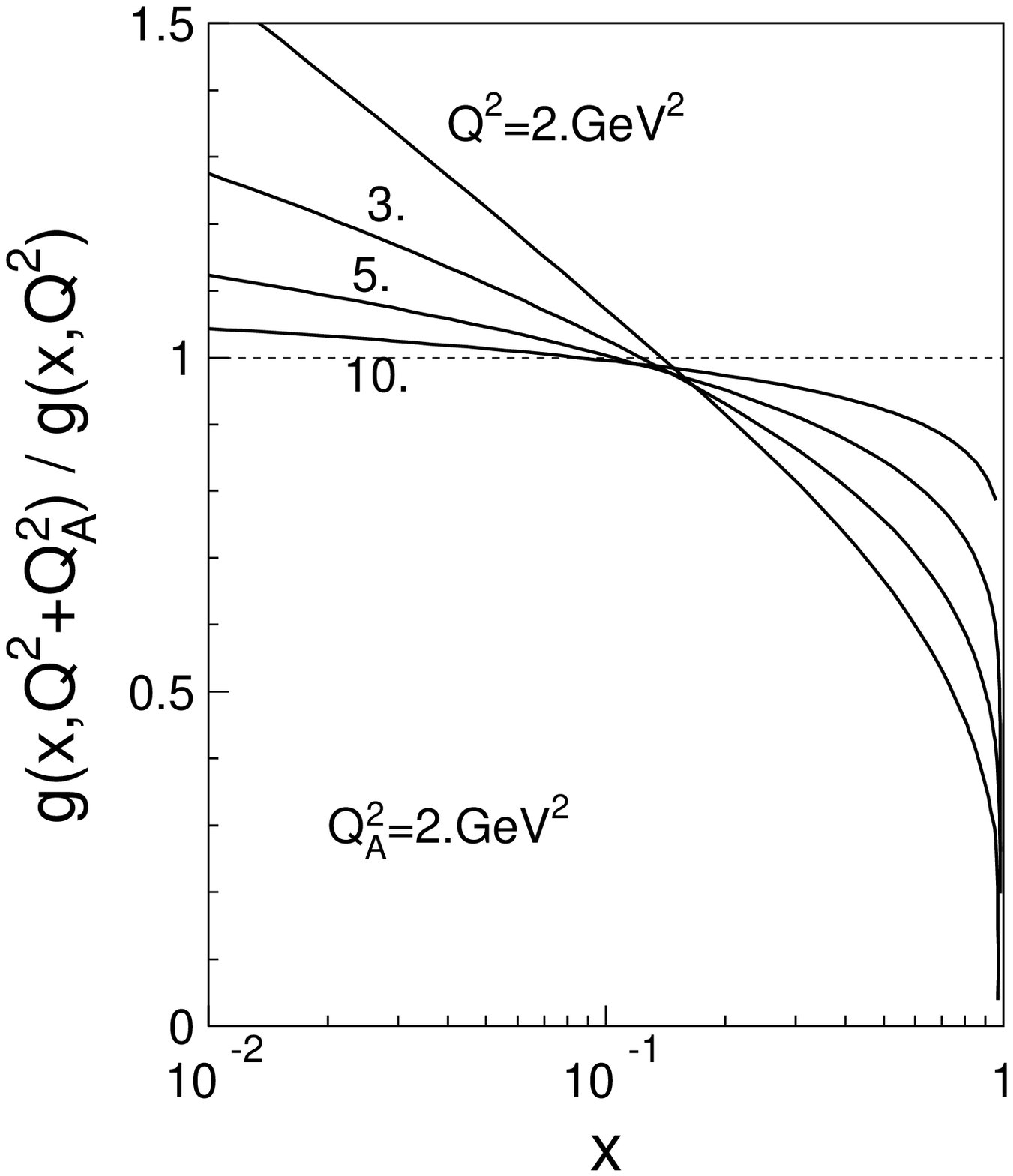}}
\caption{Ratio of the gluon distribution functions in the  proton in $pA$ and $pp$ collisions calculated with MSTW2008 \cite{mstw} . The saturation momentum $Q_{sA}^2=2\GeV^2$ and  hard scale $Q^2=2,\ 3,\ 5,\ 10\GeV^2$.}
\label{fig:rescaling}}
\hfill
\parbox{\halftext}{
\centerline{\includegraphics[width=6 cm]{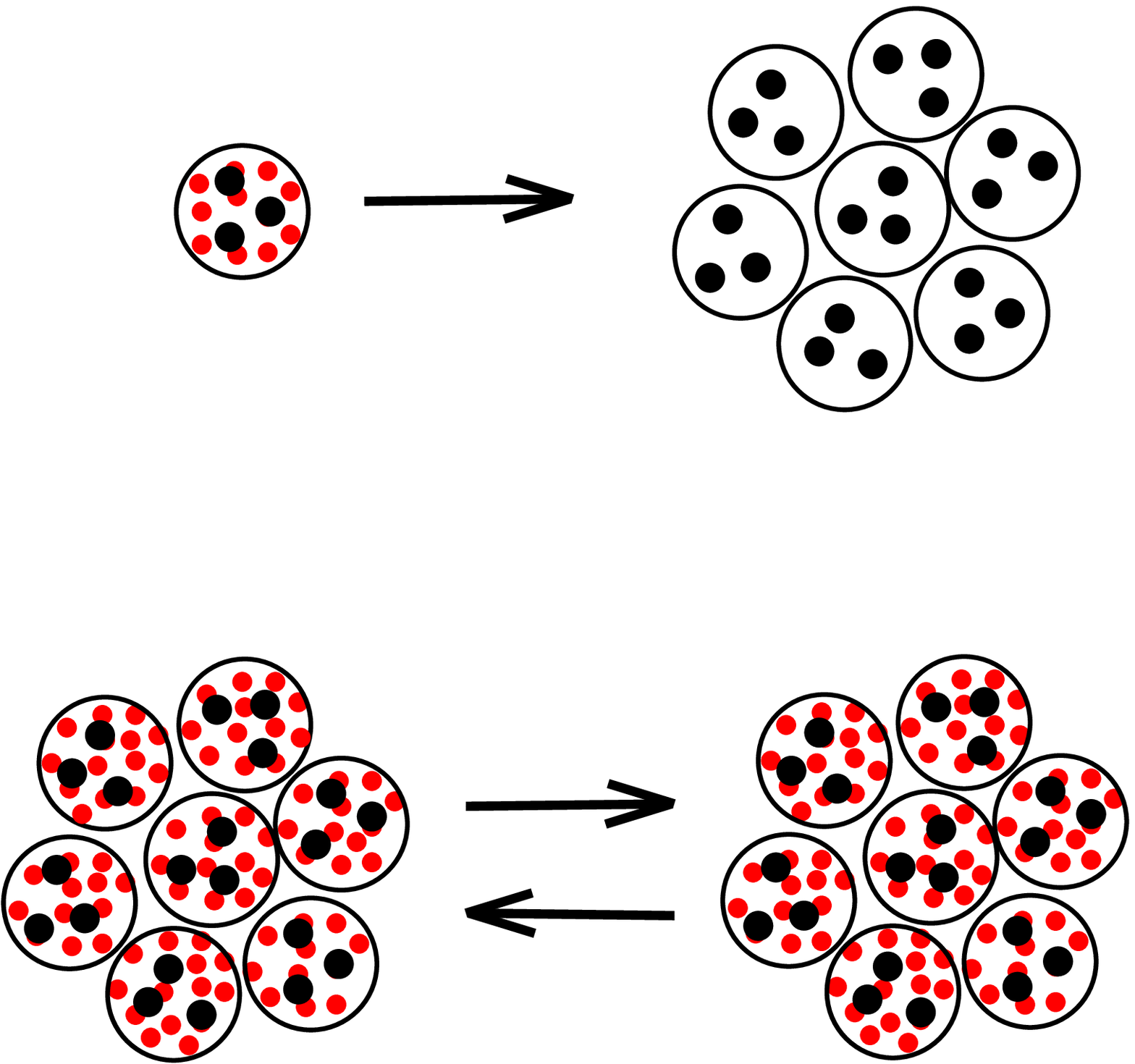}}
\caption{{\it Top:}  $pA$ collision in which the
colliding proton is excited by multiple interactions to higher Fock states which have more small-$x$ gluons. {\it Bottom:} nuclear
collision in which participating nucleons on both sides are boosted
to a higher saturation scale.}\label{fig:p-A-A}}
\end{figure}

Due to the effect of broadening a nuclear target probes the parton distribution in the beam hadron with a higher resolution, compared to a proton target, so in a hard reaction the effective scale {$ Q^2$} for the beam PDF drifts towards higher values, {$Q^2\Rightarrow Q^2+Q_{sA}^2$}
\cite{boosting}, where $Q_{sA}$ is the saturation momentum in the nucleus\footnote{We assume that the coherence time of gluon radiation substantially exceeds the nuclear size.}.   
The projectile PDF is modified due to the selection of higher Fock states by multiple interactions.
The modified gluon distribution turns out to be suppressed at large $ x\to1$ \cite{qv}, but enhanced at small $ x$. This is a higher twist effect.
Examples of $pA$ to $pp$ ratios of the gluon densities in the beam proton, $g(x,Q^2+Q_{sA}^2)/g(x,Q^2)$, are presented in Fig.~\ref{fig:rescaling} for a hard
reaction (high-$p_T$, heavy flavor production, etc.) at different hard scales. 

Apparently, there is an asymmetry in the properties of colliding nucleons in  {$ pA$} collisions:
the PDF of the beam proton is modified to a state with a higher hard scale and a larger gluon density  at small {$ x$} than in {$ NN$} collisions. At the same time, the properties of the target  bound nucleons remain unchanged, because they do not undergo multiple interactions. This feature is illustrated pictorially in the upper part of Fig.~\ref{fig:p-A-A}. The picture in the bottom part of this figure demonstrates that in the case of a nuclear collision, the PDFs of bound nucleons in both nuclei are drifting towards higher parton multiplicities. The new scales depend on each other: the higher is the saturation scale in the nucleus $A$, the more is the shift of the saturation scale in the nucleus $B$, and vice versa.

Such a mutual influence of the scales can be described by bootstrap equations~\cite{boosting},
 \beqn
 \label{430}
\tilde Q_{sB}^2(x_B)&=&\frac{3\pi^2}{2}\,\alpha_s(\tilde Q_{sA}^2+Q_0^2)\,
x_B g_N(x_B,\tilde Q_{sA}^2+Q_0^2)\,T_B;
\nonumber\\
\tilde Q_{sA}^2(x_A)&=&\frac{3\pi^2}{2}\,\alpha_s(\tilde Q_{sB}^2+Q_0^2)\,
x_Ag_N(x_A,\tilde Q_{sB}^2+Q_0^2)\,T_A.
\eeqn
Here
{$ x_{A,B}$} are the fractional light-cone momenta of the radiated gluon relative to the colliding nuclei, {$ x_Ax_B= k_T^2/s$}; {$ Q_0^2=1.7\GeV^2$} is chosen to get the correct infra-red behavior.
These equations lead to a larger nuclear saturation scale $\tilde Q_{sA}^2$ in $AA$ compared to $Q_{sA}^2$ in $pA$ collisions. 

\begin{figure}[htb]
\parbox{\halftext}{
\centerline{\includegraphics[width=4.5 cm]{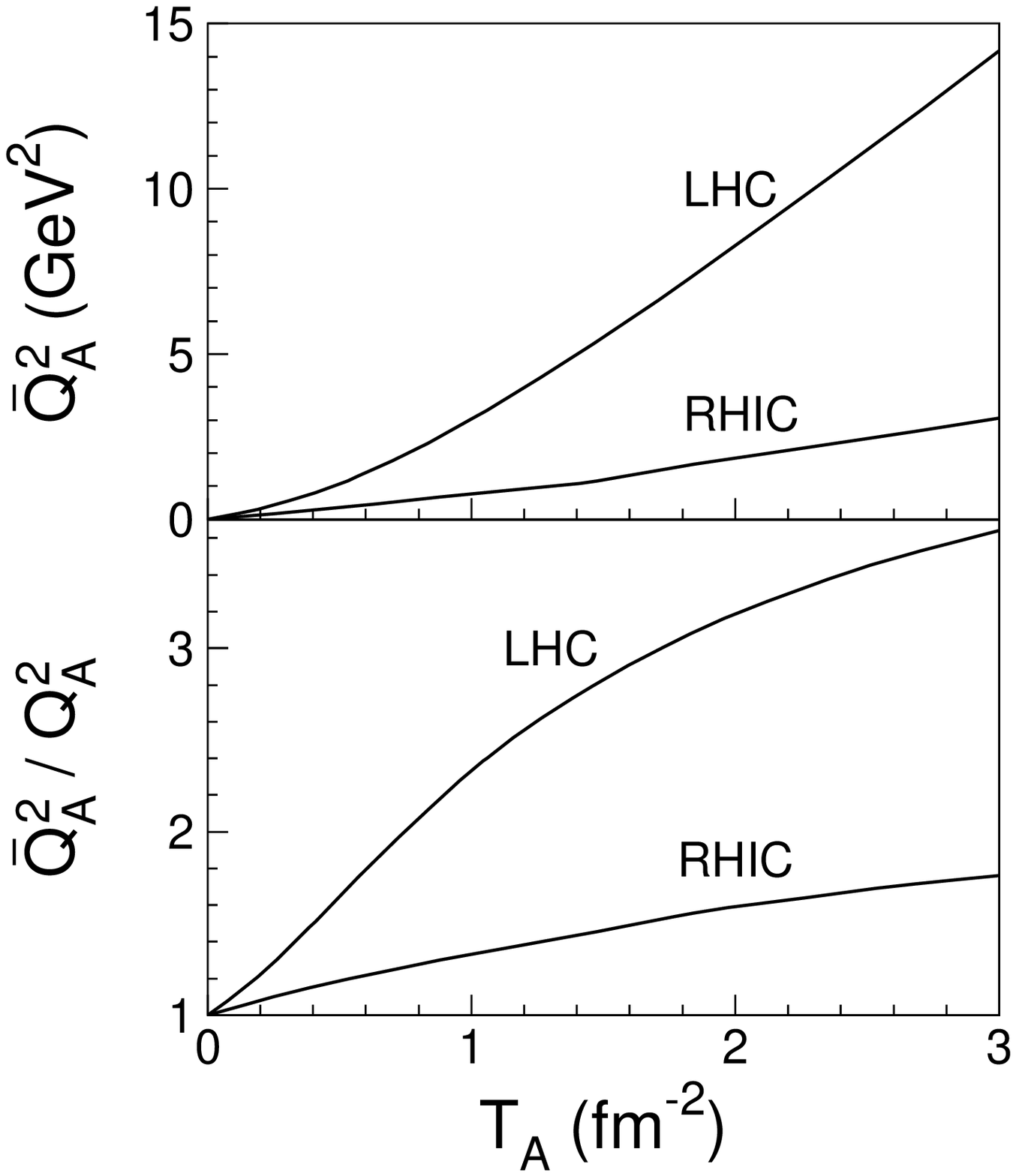}}
\caption{{\it Top panel:}  The boosted values of the saturation momentum 
squared $\tilde Q_{sA}^2$ calculated at the energies of RHIC and LHC as function of nuclear thickness $T_A=T_B$.
 {\it Bottom panel:} the ratio $\tilde Q_{sA}^2/Q_{sA}^2$ as function of $T_A=T_B$.}\label{fig:boosting}}
\hfill
\parbox{\halftext}{
\centerline{\includegraphics[width=4 cm]{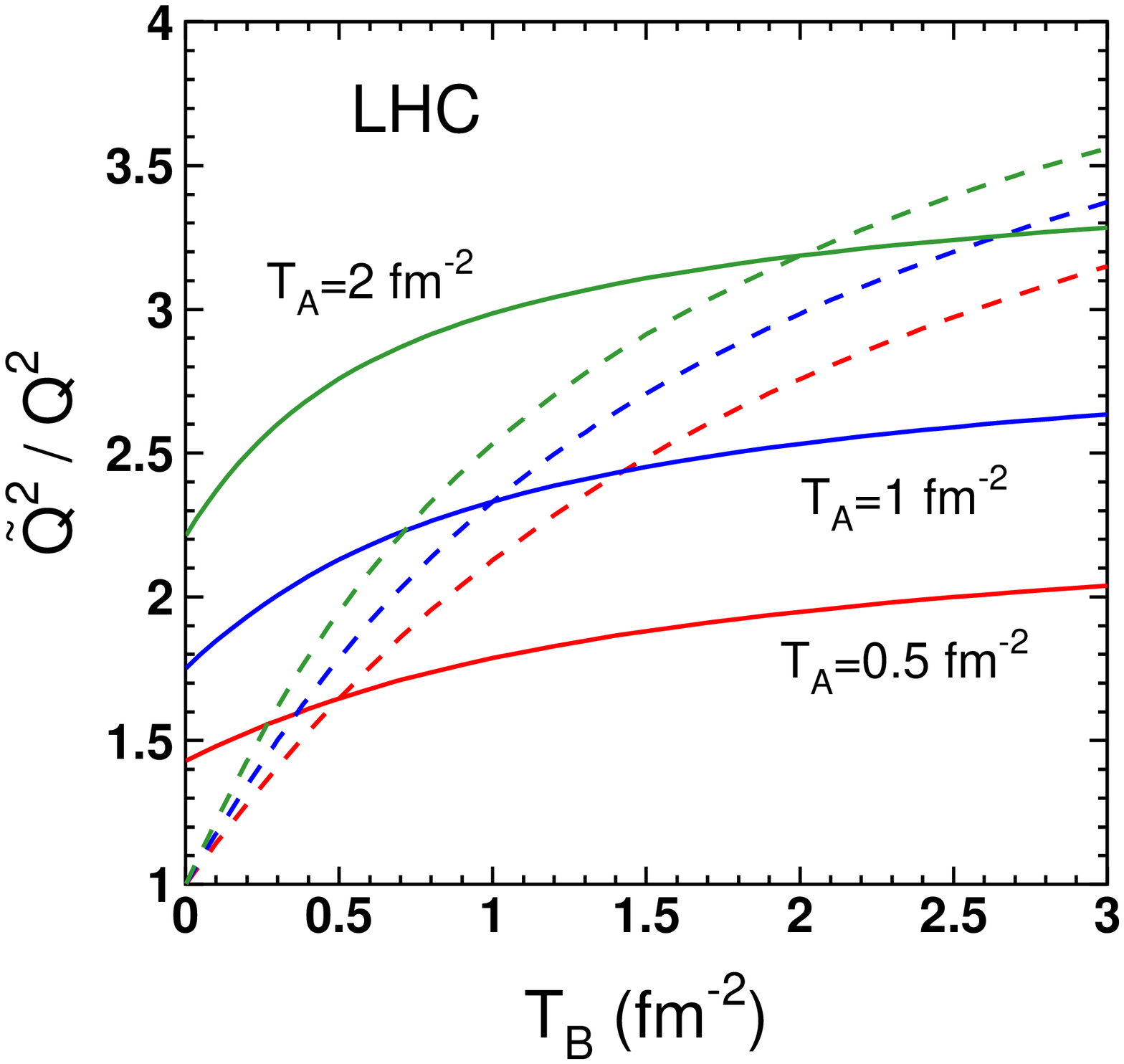}}
\caption{Solutions of the reciprocity equations (\ref{430}) for non-central collision of identical nuclei $A$ and $B$ at $x_A=x_B=0$.
Dashed curves correspond to $\tilde Q_{sA}^2/Q_{sA}^2$, the solid curve to $\tilde Q_{sB}^2/Q_{sB}^2$ as function of $T_B$. In both cases $T_A$ is fixed at $0.5$, $1.0$ and $2.0\fm^{-2}$ from bottom to top.
}\label{fig:TA-TB}}
\end{figure}

The solution of equations (\ref{430}) for $\tilde Q_{sA}^2$ in a central  lead-lead collision at the mid rapidity is plotted as function of nuclear thickness   $T_A=T_B$ in the upper panel of Fig.~\ref{fig:boosting} at the energies of RHIC and LHC, and the ratio $\tilde Q_{sA}^2/Q_{sA}^2$ is shown in the bottom panel.
While the magnitude of the boosting effect is moderately large, about $20\%$ at the energy of RHIC, it becomes significan at LHC.

In non-central collisions $T_A\neq T_B$ and the solution of equations (\ref{430})
is presented in Fig.~\ref{fig:TA-TB} separately for $\tilde Q_{sA}^2/Q_{sA}^2$ (solid curves) and $\tilde Q_{sB}^2/Q_{sB}^2$ (dashed curves).  For each set of curves $T_A$ is fixed at $0.5$, $1$ and $2\fm^{-2}$ from bottom to top.

\section{Observables for the boosting effect}

\begin{figure}[htb]
\parbox{\halftext}{
\centerline{\includegraphics[width=4.7 cm]{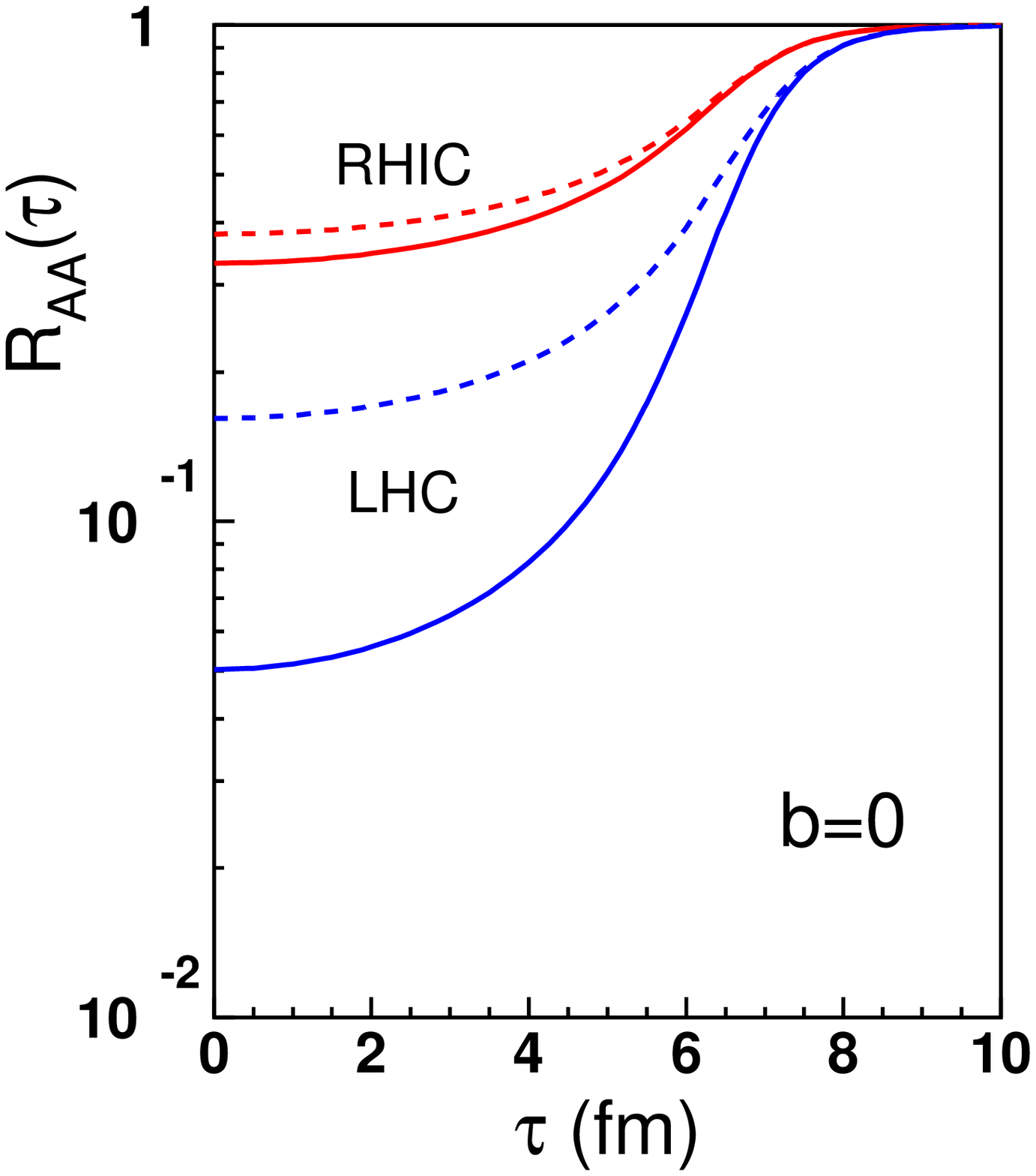}}
\caption{Initial state suppression of $J/\Psi$ produced in central $Au$-$Au$ collisions at $y=0$ as function of impact parameter $\tau$. The upper and bottom sets of curves correspond to $\sqrt{s}=200\GeV$ and $5.5\TeV$ respectively, and are calculated excluding (dashed) and including (solid) the boosting effect.}\label{fig:RAA-tau}}
\hfill
\parbox{\halftext}{
\centerline{\includegraphics[width=4.3 cm]{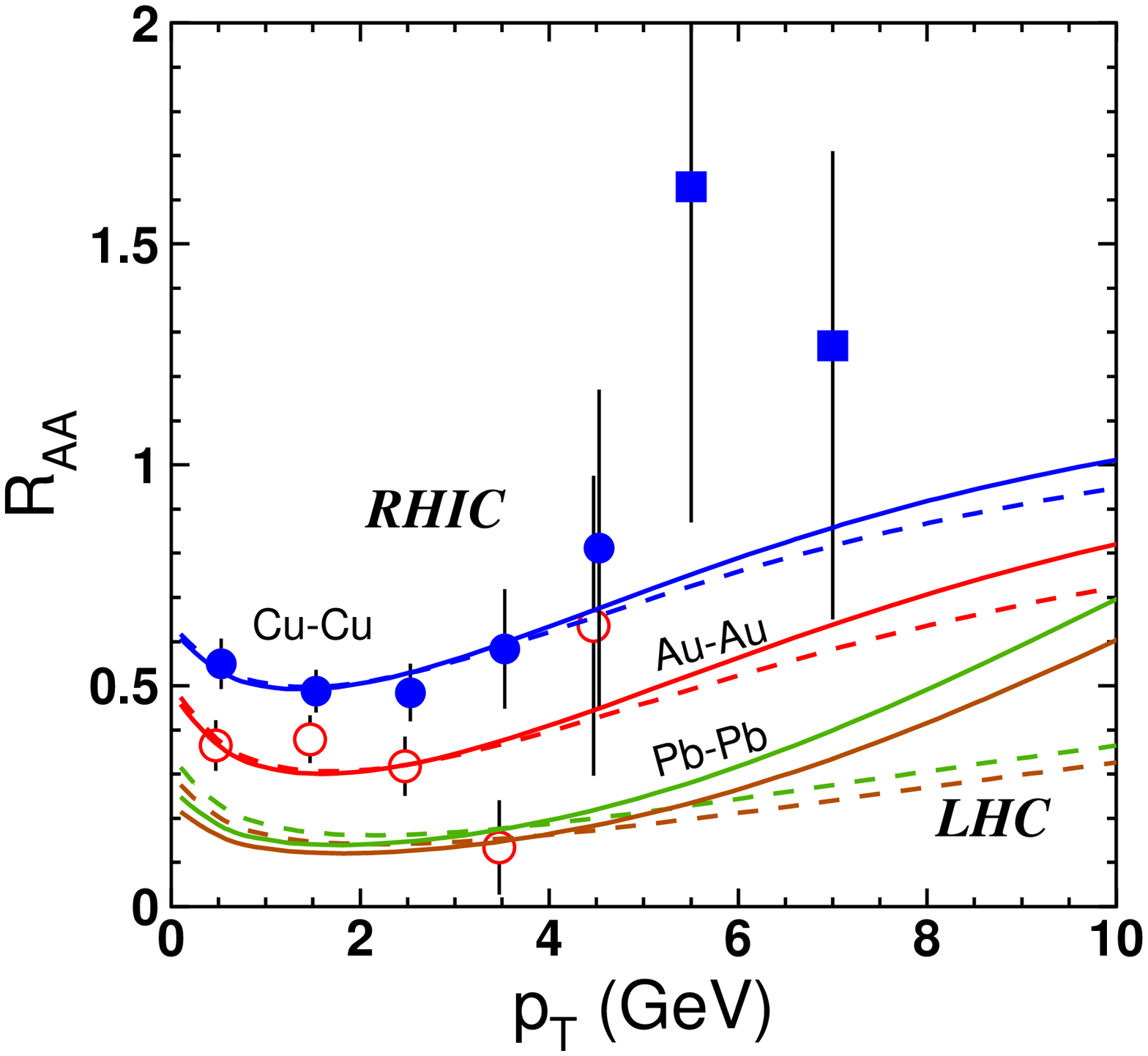}}
\caption{ Data \cite{phenix1,phenix2,star} for the nuclear modification factor $R_{AA}$ for $J/\Psi$ production in central $Cu$-$Cu$ collisions (closed points) and $Au$-$Au$ (open circles) at $y=0$ and $\sqrt{s}=200\GeV$ (RHIC). The bottom set of curves show
$R_{AA}(p_T)$ predicted at $\sqrt{s}=5.5\TeV$ (LHC). Dashed and solid curve are calculated excluding, or including the boosting effect, respectively.}\label{fig:AA-pt}}
\end{figure}

Since the nuclear medium in $AA$ collisions has an enriched gluon density at small $x$,
it is more opaque for color dipoles than what one could expect extrapolating from $pA$.
In particular, charmonium suppression by initial state interactions (ISI) should be stronger
\cite{nontrivial}.
The modified  dipole-nucleon cross section is also subject to the boosting effect,
\beq
\tilde\sigma_{dip}(r_T)=
\frac{\tilde Q_{sA}^2}{Q_{sA}^2}\,\sigma_{dip}(r_T).
\label{20}
\eeq
The magnitude of the ISI suppression of $J/\Psi$ in central $Pb$-$Pb$ collisions
as function of impact parameter $\tau$ is presented in Fig.~\ref{fig:RAA-tau}
by dashed (no boosting) and solid (magnified by boosting) curves.
Again, the boosting effect is rather small at RHIC, but is significant at the energy of LHC.

Unfortunately, most of the effects related to ISI are masked by final state interactions (FSI) of the $J/\Psi$ with the created dense medium. So it is very difficult to approve or disprove the boosting effect at the energies of RHIC, at least with the current precision of $J/\Psi$ data.
However, the large magnitude of the boosting effect expected at the energies of LHC,
make it plausible to single out the effect from $J/\Psi$ data. As far as the broadening of the mean transverse momentum squared, which is given by the saturation momentum \cite{broadening}, increases from 
$Q_{sA}^2$ to $\tilde Q_{sA}^2$, the Cronin effect should become much more pronounced.
This is demonstrated in Fig.~\ref{fig:AA-pt} on the example of the $p_T$ dependent nuclear ratio $R_{AA}(p_T)$
predicted for $J/\Psi$ produced in gold-gold collisions at RHIC~\cite{psi-AA}, and in lead-lead at LHC \cite{psi-lhc}. The result of calculation either with, or without the boosting effect agree with data \cite{phenix1,phenix2,star} at $\sqrt{s}=200\GeV$. The suppression of $J/\Psi$ by FSI is calculated with the transport coefficient
$\hat q_0=0.6\GeV^2/\fm$, which was adjusted \cite{psi-AA,psi-lhc} to describe the data.
At $\sqrt{s}=2.7\TeV $ the nuclear ratio for lead-lead is predicted using the value of the transport coefficient $\hat q_0=0.8\GeV^2/\fm$ found in the analysis \cite{high-pt} of data
on suppression of high-$p_T$ hadrons.  
The boosting effect at $p_T>5\GeV$  is strong, so may be seen even 
with not a very high statistics.

Another sensitive probe for the boosting effect would be an observation of different magnitudes of broadening of $J/\Psi$ in $pA$ and $AA$ collisions at the same path length in the nuclear matter \cite{boosting}. Indeed, broadening is related to the dipole cross section \cite{jkt}, which is enhanced by the boosting effect, Eq.~(\ref{20}). 
While RHIC data are not able so far to discriminate this weak effect, this task looks solvable at LHC.

Also the hadron multiplicity is expected \cite{kl} to correlate with the saturation scale, so
we expect a mismatch between the hadron multiplicities measured at the mid rapidity
as function of centrality. Indeed, an indication at such a break was observed at RHIC \cite{break},
but the effect is rather small, more precise data are required.

\section*{Acknowledgements}
I am grateful to Hans-J\"urgen Pirner, Irina Potashnikova and Ivan Schmidt for 
our longstanding and fruitful collaboration. I also thank the Galileo Galilei Institute for Theoretical Physics for the hospitality and the INFN for the partial support during the completion of this work.
This work was supported in part by Fondecyt (Chile) grant
1090291,  and by Conicyt-DFG grant No. 084-2009.

%


\begin{thebibliography}{99}
  
\bibitem{boosting}
  B.~Z.~Kopeliovich, H.~J.~Pirner, I.~K.~Potashnikova, I.~Schmidt,
  Phys.\ Lett.\  B{\bf 697 } (2011),  333.

\bibitem{qv}
  J.-W.~Qiu, I.~Vitev,
  Phys.\ Lett.\  {\bf B632 } (2006),  507.


\bibitem{mstw}
  A.~D.~Martin, W.~J.~Stirling, R.~S.~Thorne and G.~Watt,
  Eur.\ Phys.\ J.\  C {\bf 64} (2009), 653.

\bibitem{nontrivial}
  B.~Z.~Kopeliovich, I.~K.~Potashnikova, H.~J.~Pirner and I.~Schmidt,
  Phys.\ Rev.\  {\bf C83 } (2011), 014912.

\bibitem{broadening}
  B.~Z.~Kopeliovich, I.~K.~Potashnikova and I.~Schmidt,
  Phys.\ Rev.\  C {\bf 81} (2010), 035204.

\bibitem{psi-AA}
  B.~Z.~Kopeliovich, I.~K.~Potashnikova and I.~Schmidt,
  Phys.\ Rev.\  C {\bf 82} (2010), 024901.

\bibitem{psi-lhc}	
B.Z. Kopeliovich, I.K. Potashnikova, Ivan Schmidt
Nucl. Phys. A{\bf 864} (2011), 203. 

  \bibitem{phenix1}
  A.~Adare {\it et al.}  [PHENIX Collaboration],
  Phys.\ Rev.\ Lett.\  {\bf 98} (2007), 232301.

\bibitem{phenix2}
  A.~Adare {\it et al.}  [PHENIX Collaboration],
  Phys.\ Rev.\ Lett.\  {\bf 101} (2008), 122301.

\bibitem{star}
  B.~I.~Abelev {\it et al.}  [STAR Collaboration],
  Phys.\ Rev.\  C {\bf 80} (2009), 041902.

\bibitem{high-pt}
  B.~Z.~Kopeliovich, I.~K.~Potashnikova and I.~Schmidt,
  Phys.\ Rev.\  C {\bf 83} (2011), 021901.

\bibitem{jkt}
  M.~B.~Johnson, B.~Z.~Kopeliovich and A.~V.~Tarasov,
  Phys.\ Rev.\  C {\bf 63} (2001), 035203.

\bibitem{kl}
  D.~Kharzeev, E.~Levin, M.~Nardi,
  Nucl.\ Phys.\  {\bf A747 } (2005),  609.
  
  \bibitem{break}
  B.~Alver {\it et al.},
  Phys.\ Rev.\ Lett.\  {\bf 102} (2009) 142301.


\end{thebibliography}
\end{document}